\documentclass[submission,copyright,creativecommons]{eptcs}
\providecommand{\event}{ICLP 2026} % Name of the event you are submitting to

\usepackage{iftex}

\usepackage{graphicx}
\usepackage{amsmath}
\usepackage{amsthm}
\newtheorem{definition}{Definition}[section]

\usepackage{listings}
\ifpdf
  \usepackage{underscore}         % Only needed if you use pdflatex.
  \usepackage[T1]{fontenc}        % Recommended with pdflatex
\else
  \usepackage{breakurl}           % Not needed if you use pdflatex only.
\fi

\title{CaVE: A Constraint Storage Approach to Handling Integrity Constraints}
\author{Xiangyu Guo \quad\qquad Ajay Bansal
\institute{School of Computing and Augmented Intelligence\\
Arizona State University, Tempe, Arizona}
\email{Xiangyu.Guo@asu.edu \quad\qquad Ajay.Bansal@asu.edu}
}
\def\titlerunning{CaVE: A Constraint Storage Approach to Handling Integrity Constraints}
\def\authorrunning{Xiangyu Guo and Ajay Bansal}
\begin{document}
\maketitle

\begin{abstract}
This paper presents Constraints as Verifiers and Emitters (CaVE), a constraint storage approach for handling integrity constraints in stableKanren.
stableKanren is a normal logic-program solver based on extended unification and resolution.
Integrity constraints control the outcomes of goals in normal logic programs, which is critical for non-monotonic reasoning.
There is no resolution-based algorithm for handling integrity constraints that can be used in stableKanren.
Therefore, we design Constraints as Verifiers and Emitters (CaVE), a constraint storage that works with resolution to support integrity constraints.
We discuss variants of CaVE with respect to the integrity constraints, ranging from propositional to grounded and predicate versions.
We implement CaVE using Scheme in stableKanren and show a series of example normal programs written in stableKanren with integrity constraints.
\end{abstract}

\section{Introduction}
\label{sec:intro}
    In this paper, we propose a novel approach for handling integrity constraints that is compatible with resolution-based logic-program solvers.
    A normal logic program (Definition \ref{def:normal-program}) under stable model semantics \cite{Gelfond:1988:stable} is capable of non-monotonic reasoning, an ability to reason about contradictions.
    For example, Alice and Bob are planning a business trip, but only one person can travel, either Alice or Bob.
    A propositional normal program to create a travel plan written in Prolog syntax \cite{ISO:1995:IIIe} as follows.
    \begin{lstlisting}
alice :- not bob.
bob :- not alice.
    \end{lstlisting}
    The example program forms a \textit{loop}, where the truth value of an atom depends on itself.
    This loop also involves negations (\textit{not}), so it is a \textit{negative loop}.
    The number of negations in the loop is an even number; therefore, it is an \textit{even negative loop}.
    According to stable model semantics, the example program has two models, $\{alice\}$ and $\{bob\}$.
    Either \textit{alice} or \textit{bob} can be true, but they cannot be true or false at the same time.
    Therefore, either Alice or Bob can travel, but they cannot travel together.

    Lin and Zhao \cite{Lin:2004:odd-and-even-cycles} point out that even negative loops in the normal program generate combinations, and the odd negative loops eliminate combinations.
    The integrity constraint has an alternative representation (Definition \ref{def:alter-constraint-rule}) using an odd negative loop to eliminate models.
    For instance, an integrity constraint to prevent Alice from traveling is as follows.
    \begin{lstlisting}
fail :- alice, not fail.
    \end{lstlisting}
    The odd negative loop forms a contradiction if \textit{alice} is true.
    A simpler headless equivalent integrity constraint is written as follows.
    \begin{lstlisting}
:- alice.
    \end{lstlisting}
    Adding this integrity constraint to the example program produces only one model, $\{bob\}$.
    Section \ref{sec:preliminaries} provides a detailed review of the integrity constraint.
    
    stableKanren extends miniKanren, a relational logic program solver that implements resolution and unification \cite{Friedman:2005:ReasonedSchemer1st}, with stable model semantics to handle the negation rule in normal logic programs \cite{Guo:2023:PPDP-stableKanren}.
    This extension allows stableKanren to reason about contradictions.
    The same travel planning program is written in stableKanren syntax as follows.
    \begin{lstlisting}
(defineo (alice) (noto (bob)))
(defineo (bob) (noto (alice)))
    \end{lstlisting}
    The stableKanren syntax is similar to the functional programming language Scheme \footnote{https://cisco.github.io/ChezScheme/}.
    It uses \textit{defineo} to create goal functions or logic programming clauses and uses \textit{noto} for negation.
    The goal function name is the head of the normal logic program clause.
    The goal function body is the body of the normal logic program clause.
    Section \ref{sec:mk-sk} gives a complete introduction to miniKanren and stableKanren.

    A propositional integrity constraint to eliminate models in stableKanren is as follows.
    \begin{lstlisting}
(defineo (fail) (alice) (noto (fail)))
    \end{lstlisting}
    However, the predicate integrity constraint written in this form does not work as expected under resolution.
    Supporting predicate integrity constraints using resolution is not straightforward.
    The main reason is that the resolution tends to leave the variables in the predicate integrity constraints unbound.
    The resolution's minimal model trait ensures that it binds the minimum number of variables or proves the minimum number of sub-goals to be true.
    The body variables in the integrity constraints that remain unbound can satisfy the integrity constraint under resolution.
    Therefore, the predicate integrity constraints under resolution do not work as expected.
    To address the issue, existing algorithms either employ grounding or necessitate additional modifications to the resolution process.
    The grounding approach does not align with the top-down resolution, and the modified resolution cannot terminate the resolution branch once a violated constraint is encountered.
    Section \ref{sec:bottom-up-top-down} delivers a comprehensive discussion on the existing algorithms.

    We set aside integrity constraints for resolution and treat them as exceptions to the resolution.
    The resolution only processes the normal program clauses, and the integrity constraints collect the outcome of the resolution.
    The integrity constraint is verified when it has sufficient results.
    We design Constraints as Verifiers and Emitters (CaVE), a constraint storage approach, to achieve this purpose.
    Section \ref{sec:cave} presents the CaVE algorithm and discusses the different cases of CaVE.
    In Section \ref{sec:continuation}, we show that the constraint store in CaVE can be easily implemented in stableKanren using the functional programming language Scheme.
    Section \ref{sec:csp-in-sk} demonstrates examples of using CaVE in stableKanren to solve combinatorial search problems.

\section{Preliminaries}
\label{sec:preliminaries}
    This section reviews several key definitions in logic programming.
    Lloyd defines \textit{definite program}, \textit{normal program}, and \textit{integrity constraints} as follows \cite{Lloyd:1987:FoundationsOFLP}.

    \begin{definition}[definite program clause]
    \label{def:definite-program-clause}
    A \textit{definite program clause} is a clause of the form,
    \[A \leftarrow B_1, \cdots, B_n\]
    where $A, B_1, \dots , B_n$ are atoms\footnote{Atom is evaluated to be true or false.}.
    \end{definition}
    A definite program clause contains precisely one atom A in its consequent.
    $A$ is called the \textit{head} and $B_1, \dots , B_n$ is called the \textit{body} of the program clause.
    \begin{definition}[definite program]
    \label{def:definite-program}
    A \textit{definite program} is a finite set of definite program clauses.
    \end{definition}

    Based on the definition of the definite program clause, we have the definition of \textit{normal program clause} and \textit{normal program}.
    \begin{definition}[normal program clause]
    \label{def:normal-program-clause}
    A \textit{normal program clause} is a clause of the form,
    \[A \leftarrow B_1, \cdots , B_n, not \; B_{n+1}, \cdots , not \; B_{m}\]
    \end{definition}
    For a normal program clause, the body of a program clause is a conjunction of literals instead of atoms; $B_1, \cdots, B_n$ are \textit{positive literals} and $not \; B_{n+1}, \cdots, not \; B_{m}$ are \textit{negative literals}\footnote{A positive literal is just an atom, a negative literal is the negation of an atom.}.
    \begin{definition}[normal program]
    \label{def:normal-program}
    A \textit{normal program} is a finite set of normal program clauses.
    \end{definition}

    The \textit{integrity constraint} is defined as a headless normal program clause.
    \begin{definition}[integrity constraint]
    \label{def:constraint-rule}
    An \textit{integrity constraint} is a clause of the form,
    \[ \bot \leftarrow B_1, \cdots , B_n, not \; B_{n+1}, \cdots , not \; B_{m}\]
    where $B_1, \dots , B_m$ are atoms.
    \end{definition}
    For an integrity constraint, the head of a clause is $\bot$ (false) instead of an atom.
    So, the integrity constraints are headless.
    The constraint is violated only if all body literals are evaluated as $\top$ (true).
    When the clause's body is evaluated as $\top$, we get a formula that says true implies false.
    \[\bot \leftarrow \top\]
    This formula is impossible to prove true.
    So, the constraint is violated.
    The constraint can be satisfied if any literal in the body is evaluated to be false.
    Therefore, the integrity constraints mean that not all literals in the body can be proven true. The integrity constraint has an alternate form in Definition \ref{def:alter-constraint-rule}.
    \begin{definition}[alternate integrity constraint]
    \label{def:alter-constraint-rule}
    An \textit{alternate integrity constraint} is a clause of the form,
    \[ \boldsymbol{fail} \leftarrow B_1, \cdots , B_n, not \; B_{n+1}, \cdots , not \; B_{m}, \boldsymbol{not \; fail}\]
    \end{definition}
    This translation is equivalent to the original integrity constraint in Definition \ref{def:constraint-rule}; no other transformations are needed.
    The \textit{fail} and \textit{not fail} pair introduces a contradiction when all other clauses' bodies $B_i$ are evaluated as true.

\section{Related Work}
\label{sec:related}
    This section first introduces miniKanren and stableKanren syntax and query interface.
    The miniKanren programs are corresponding to definite programs (Definition \ref{def:definite-program}) and the stableKanren programs are corresponding to normal programs (Definition \ref{def:normal-program}).
    Later, we discuss the challenges of adding integrity constraints to stableKanren and justify the need for a new algorithm.

\subsection{miniKanren and stableKanren}
\label{sec:mk-sk}
    Friedman et al. build miniKanren to capture unification and resolution, the essence of Prolog, and show a natural way to extend functional programming to relational programming \cite{Friedman:2005:ReasonedSchemer1st}.
    The core miniKanren implementation \footnote{https://github.com/miniKanren/simple-miniKanren} introduces only a few operators to users: $==$ for \emph{unification}, $fresh$ for \emph{existential quantification}, $conde$ for \emph{disjunction}, and a $run$ interface for running the query.
    The \emph{conjunction} is captured by the sequential evaluation process naturally, so there is no operator for conjunction.

    The miniKanren syntax uses \textit{goal functions} to represent Prolog predicates.
    The translation from Prolog syntax to miniKanren syntax is straightforward.
    For example, a Prolog program written below,
    \begin{lstlisting}
p(X) :- X = 1. p(X) :- a(X). a(Y) :- Y = cat.
    \end{lstlisting}
    has a miniKanren equivalent.
    \begin{lstlisting}
(define (p X) (conde [(== X 1)] [(a X)]))
(define (a Y) (conde [(== Y 'cat)]))
    \end{lstlisting}
    The keyword \textit{define} defines a miniKanren goal function.
    Each head of the Prolog clause corresponds to a miniKanren \textit{goal function}, and each clause's head variables are the goal function's parameters.
    A set of Prolog clauses with the same head is grouped under a single goal function.
    Inside the goal function, \textit{conde} represents the non-deterministic choice of different clauses in Prolog.
    The miniKanren unification \textit{==} is the same as the Prolog unification \textit{=}.

    A query starts an automated resolution process, and miniKanren provides a \textit{run} interface to execute a query.
    For example, a query on the example program, along with its output, is shown as follows.
    \begin{lstlisting}
> (run 3 (V) (p V))
(1 cat)
    \end{lstlisting}
    The \textit{run} interface has three parameters.
    The first parameter is the number of answers we expect; the query returns no more than that.
    The second parameter is the query variable, which stores the query results.
    The third parameter is the actual query.
    In this example, the query returns two answers for $V$ as a list: 1 and cat.

    Guo et al. extend miniKanren under stable model semantics to support reasoning about contradiction in stableKanren \footnote{https://github.com/stable-Kanren/stable-Kanren} \cite{Guo:2023:PPDP-stableKanren}.
    In addition to miniKanren's operators, stableKanren introduces two new operators, \textit{defineo} and \textit{noto}.
    The \textit{noto} operator represents the negation (\textit{not}) in the normal program clause (Definition \ref{def:normal-program-clause}).
    stableKanren uses \textit{defineo} to define its goal functions instead of \textit{define} in miniKanren.
    In the travel planning example shown in Section \ref{sec:intro}, Alice and Bob are planning a business trip, but only one person can travel: either Alice or Bob.
%     The condition can be represented as a stableKanren program as follows.
%     \begin{lstlisting}
% (defineo (Alice)      (defineo (Bob)
%   (noto (Bob))          (noto (Alice)))
%     \end{lstlisting}
%     In the example, there is no variable in each goal function, so it is equivalent to propositional logic.
%     The success of the goal function means the person can travel, and failure means they cannot.
    The same \textit{run} interface is used for the query.
    The following set of queries and outputs is shown.
    \begin{lstlisting} 
> (run 1 (q) (bob))
(_.0)
> (run 1 (q) (alice) (bob))
()
    \end{lstlisting}
    The first query asks: ``Can Bob go travel?''
    It returns a list containing one element, ``\_.0'', which represents anything in miniKanren and stableKanren.
    Anything can let the goal function \textit{Bob} succeed.
    Hence, Bob can travel.
    The second query asks: ``Can Alice and Bob travel together?''
    It returns an empty list, representing nothing in miniKanren and stableKanren.
    In this case, neither goal function \textit{Alice} nor \textit{Bob} can succeed.
    Therefore, they cannot travel together.

\subsection{Integrity Constraint Handling Methods}
\label{sec:bottom-up-top-down}
    The integrity constraint in Definition \ref{def:constraint-rule} is headless.
    One potential way to express integrity constraints in stableKanren is to utilize the alternate form presented in Definition \ref{def:alter-constraint-rule}.
    All integrity constraints are under the same goal function \textit{fail}, and the resolution ensures that they are processed at the end of solving other goals.
    As we have shown in Section \ref{sec:intro}, using a propositional integrity constraint prevents Alice from traveling.
    However, this approach does not work for predicate integrity constraints.
    For example, consider a normal logic program for solving the nqueens problem, which places $n$ queens on an $n \times n$ chessboard such that no queen attacks another queen.
    A predicate \textit{queen(x, y)} represents the queen's position $(x, y)$ on the chessboard.
    An integrity constraint to ensure that no queens attack each other on the same row is as follows
    \begin{lstlisting}
fail :- queen(X, Y), queen(U, V), X = U, Y != V, not fail.
    \end{lstlisting}
    % The variables in the integrity constraints raise two issues.
    % Firstly, the numeric computation using the variables in the integrity constraints breaks the relational property.
    % Secondly, the resolution tends to leave the variables unbound to maintain their minimal model trait.
    The resolution tends to leave the variables in the integrity constraint unbound, thereby preserving the integrity constraint's minimal model trait.
    In the normal program clauses, resolution starts from the head, which represents the main goal, and proceeds to the body, which comprises a set of sub-goals.
    Resolution attempts to prove each sub-goal, and the process continues after all sub-goals have been proven.
    Once all sub-goals are proved, all variables are bound to some values in the updated substitution set.
    In contrast, if all sub-goals in the integrity constraints have been proven true, the resolution process is terminated.
    Therefore, the smallest effort required to resolve the integrity constraints is to leave the variables unbound by resolution.

    To resolve this issue, two algorithms have been developed.
    Gebser et al. use \emph{grounding} and \emph{constraint propagation} to obtain a model of the input program \cite{Gebser:2007:conflict-driven-answer-enumeration,Gebser:2007:conflict-driven-answer-solving}.
    The predicate integrity constraint is treated the same as other normal program clauses, so the grounding removes variables in the predicate integrity constraint \cite{Kaminski:2023:asp-grounding-foundations}.
    The predicate integrity constraint used to guarantee no queen attacks another on the same row produces 48 propositional constraints on a $4 \times 4$ chessboard after grounding.
    \begin{lstlisting}
:-queen(1,1),queen(1,2). :-queen(1,1),queen(1,3). :-queen(1,1),queen(1,4).
                      ... ... 42 more rules ... ...                                
:-queen(4,4),queen(4,1). :-queen(4,4),queen(4,2). :-queen(4,4),queen(4,3).
    \end{lstlisting}
    During the runtime, each grounded atom assigns a true or false; the solver adjusts the truth assignment until no clause is violated.
    This approach does not align with the top-down resolution used in stableKanren.
    Additionally, the grounding stage leaves a heavy memory footprint.
    There may have been unused propositional constraints during solving, but the solver still generates them.
    It also maintains all constraints, even when they are satisfied.
    This memory bottleneck becomes a significant issue when applying the solver to large-scale problems.

    Marple et al. and Joaqun et al. convert and append the predicate integrity constraint at the end of the resolution \cite{Joaqun:2018:iclp-scasp,Marple:2017:SASP}.
    So that not all constraints are generated in the beginning.
    However, the resolution still attempts to find new values for the unbound variables in the predicate integrity constraints.
    As we have discussed earlier, the resolution leaves all variables unbound to satisfy the integrity constraints.
    This does not align with the integrity constraint's purpose, which requires variables to be bound to specific values for verification.
    % Internally, it uses \textit{NMR\_CHECK transformation} \cite{Marple:2017:SASP} to convert each integrity constraint into nested \textit{forall} so that not all constraints are generated at the beginning, but after resolution.
    For example, an equivalent of nqueens row checking is written in s(CASP) syntax as follows.
    \begin{lstlisting}
:- use_module(library(scasp)). num(1). num(2). num(3). num(4).
queen(X, Y) :- num(X), num(Y), not free(X, Y).
free(X, Y) :- num(X), num(Y), not queen(X, Y).
false :-  queen(X, Y1), queen(X, Y2), Y1 \= Y2.
    \end{lstlisting}
    A query on the above program using s(CASP) \footnote{\url{https://swish.swi-prolog.org/example/scasp.swinb}},
    \begin{lstlisting}
?- queen(1, 1), queen(1, 2).
    \end{lstlisting}
    produces a wrong answer, saying $(1, 1)$ and $(1, 2)$ can be picked at the same time, but the constraint does not allow this combination.
    Hence, this approach does not properly handle predicate integrity constraints.

    We can modify to properly handle the appended integrity constraints.
    However, this generate-and-test approach still waits until the very last moment to determine if a constraint is violated, even if the constraint could terminate the computation early.
    We want to propagate the constraint as early as possible to save unnecessary computation.
    But we also do not want to ground all pairs of constraints initially.
    
    There is another issue that a logic programming solver needs to handle: operations other than unification, such as numeric computations.
    For example, an integrity constraint to ensure that no queens attack each other on the diagonal is as follows.
    \begin{lstlisting}
fail :- queen(X, Y), queen(U, V), X != U, Y != V, abs(X - U) = abs(Y - V), not fail.
    \end{lstlisting}
    For the bottom-up solver, the grounding also handles these numeric computations by generating all possible outcomes.
    For the top-down solver, there are two approaches: an impure operator or a constraint store.
    Allowing an impure operator like ``is'' in the program breaks the relational property in a logic programming language.
    Therefore, constraint logic programming (CLP) \cite{Jaffar:1987:Constraint-Logic-Programming} uses an external constraint storage to support numeric constraints \cite{Alvis:2011:ckanren,Hemann:2017:microKanrenConstraints}.
    In CLP, the solver constructs a constraint storage to track all numeric constraints during the resolution process.
    To support integrity constraints and computations other than unification without modifying the resolution, we design a constraint storage approach that tracks and updates integrity constraints alongside the resolution.

\section{Constraints as Verifiers and Emitters (CaVE)}
\label{sec:cave}
    Our algorithm treats integrity constraints as exceptions during resolution, and violations of these exceptions prune the resolution branch.
    We divide the atoms $B_1 \cdots B_m$ in an integrity constraint into two components: an \textit{emitter} and a \textit{verifier}, and represented as a constraint handler.
    \begin{definition}[constraint handler]
    \label{def:constraint-handler}
        An \textit{integrity constraint} is a constraint handler of the form,
    \begin{align}
        \label{eq:emitter}
        \bot \leftarrow E_{1}, \cdots , E_{n}, not \; E_{n+1}, \cdots , not \; E_{m}, \\
        \label{eq:verifer}
    V_{1}, \cdots, V_{i}, not \; V_{i+1}, \cdots , not \; V_{j}
    \end{align}
    where $E_1, \dots, E_m$ are emitters,
    $V_1$, \dots, $V_j$ are verifiers, and the body of the clause is a constraint handler.
    \end{definition}
    \begin{definition}[emitter]
        An \textit{emitter} is the head atom in a normal program clause.
    \end{definition}    
    Two kinds of values are emitted from the emitter: the truth value of the literal and the symbols bound to the variables of the literal.
    When a resolution successfully proves a literal in the normal program clause, the proved literal emits both kinds of values.
    The emitter can be further categorized into two types: \textit{constant emitter} and \textit{variable emitter}.
    \begin{definition} [constant emitter]
    \label{def:constant-emitter}
        A constant emitter is an emitter that only emits a truth value.
    \end{definition}
    \begin{definition} [variable emitter]
    \label{def:variable-emitter}
        A variable emitter is an emitter that not only emits a truth value, but also emits symbols to unique variables.
    \end{definition}
    The truth value is directly emitted to the constraint handler (Equation \ref{eq:emitter} in Definition \ref{def:constraint-handler}).
    The symbols are emitted to the verifiers through unique variables.
    \begin{definition} [verifier]
        A \textit{verifier} is a boolean expression with unique variables that receives and verifies symbols from the emitter.
    \end{definition}
    All verifiers are also combined into a constraint handler (Equation \ref{eq:verifer} in Definition \ref{def:constraint-handler}). 
    Therefore, if all literals are proved to be true and all verifiers are true, the constraint handler returns true, indicating that an integrity constraint has been violated.
    
    \begin{definition} [CaVE]
    \label{theorem:CaVE}
        An integrity constraint breaks into emitters and verifiers, forming a constraint handler.
        The emitters emit values (truth value and symbols) during resolution.
        The verifiers receive symbols from the emitters.
        The constraint handler receives truth values from the emitters and the evaluation outcome of verifiers.
        Once all emitters have emitted values, the verifiers have collected a sufficient number of symbols.
        The constraint handler then verifies the constraint, and the verification result controls the resolution.
    \end{definition}
    As we mentioned in Section \ref{sec:preliminaries}, the integrity constraints eliminate only the models.
    Therefore, we need to show that CaVE eliminates models that violate the integrity constraints.
    To begin with, we first show the propositional case, which is simpler without any verifiers and only has constant emitters.
    Once the variables are introduced into the integrity constraint, not all body atoms $B_i$ are emitters; some are verifiers.
    We discuss how to categorize the atoms in the integrity constraints into emitters and verifiers after Definition \ref{corollary:grounded-CaVE}.
    \begin{definition} [Propositional CaVE]
    \label{lemma:CaVE}
        Given a propositional program, the propositional integrity constraint in such a program only has constant emitters.
        Constant emitters emit truth values only during the resolution process.
        The constraint handler only collects truth values from constant emitters.
    \end{definition}

    % \begin{proof}
        For a propositional integrity constraint.
        Each atom $B_i$ in Definition \ref{def:constraint-rule} has no terms, only predicates.
        Therefore, each $B_i$ corresponds to a constant emitter $E_i$.
        The constraint handler has the exact same form as the body of the integrity constraint.
        When all body literals are evaluated as $\top$, the constraint handler returns true, and the resolution is terminated.
        When any body literal is evaluated as $\bot$, the constraint handler returns false, and the resolution continues.
        Hence, the constraint handler in propositional CaVE behaves the same as the propositional integrity constraint.
    % \end{proof}

    Using Definition \ref{lemma:CaVE}, we derive two special cases regarding the relationships between the predicate CaVE and the propositional CaVE.

    \begin{definition} [Constant CaVE]
    \label{corollary:constant-CaVE}
        An integrity constraint that only has constant emitters is propositional.
    \end{definition}
    %\begin{proof}
        For a constant integrity constraint.
        Each atom $B_i$ in Definition \ref{def:constraint-rule} has predicates and terms, but the terms are only constants.
        Therefore, each $B_i$ corresponds to a constant emitter $E_i$.
        The constraint handler has the exact same form as the body of the integrity constraint.
        When all body literals are evaluated as $\top$, the constraint handler returns true, and the resolution is terminated.
        When any body literal is evaluated as $\bot$, the constraint handler returns false, and the resolution continues.
        Hence, the constraint handler in constant CaVE behaves the same as the constant integrity constraint.
    %\end{proof}

    As we have shown so far, all body atoms $B_i$ in Definition \ref{def:constraint-rule} are corresponded to constant emitters when there are no variables present.
    Once the variables are introduced into the integrity constraint, not all body atoms $B_i$ are variable emitters; some are verifiers.
    Advancing from propositional to predicate integrity constraint, we need to distinguish verifiers from emitters in the body atoms $B_i$.
    Emitters and verifiers are connected through unique variables.
    One emitter can emit to multiple verifiers, and one verifier can receive from multiple emitters, as long as each variable in the emitters is unique.
    % We assume the variables in the emitters are unique for now, and we will relax this assumption later.
    Therefore, $B_i$ is an emitter if and only if $B_i$ is the head atom in a normal program clause; otherwise, $B_i$ is a verifier, a boolean expression with variables that receives and verifies symbols from the emitter.
    When the emitters use the same variable $V_i$ more than once, the variables in the emitter are not unique.
    This means that the values emitted by these emitters must match the integrity constraint.
    We convert this variable $V_i$ to unique variables by introducing a suffix numbering to each appearance of the $V_i$ from $V_{i0}$ to $V_{in}$ in the emitters and an additional verifier $V_{i0} = V_{i1} = \dots = V_{in}$ to ensure all values are the same.
    We treat all emitters' variables as unique to simplify future discussion.
    
    \begin{definition} [Grounded CaVE]
    \label{corollary:grounded-CaVE}
        An integrity constraint with variable emitters can be converted into multiple propositional integrity constraints through grounding.
    \end{definition}
    % \begin{proof}
        For an integrity constraint.
        Some atoms $B_i$ in Definition \ref{def:constraint-rule} have predicates and terms, and the terms have unique variables.
        The domain of the variables is finite; otherwise, the problem is undecidable, we are unsure whether any value violates the constraint.
        All variables can be grounded in a combination of values; the total number of combinations $S$ is the Cartesian product of the domain size of each variable.
        Therefore, all variable emitters are converted to $S$ combinations of constant emitters and create $S$ constraint handlers.
        In these grounded constraint handlers, all the verifiers can be evaluated to $\top$ or $\bot$ after grounding.
        When the verifiers are evaluated as $\bot$, the constraint handler returns false.
        When the verifiers are evaluated as $\top$, the constraint handler is pending the truth values from the constant emitters.
        The latter case we have shown in Definition \ref{corollary:constant-CaVE}.
        Hence, the multiple constraint handlers in grounded CaVE work the same as the integrity constraint.
    % \end{proof}

    Definition \ref{corollary:constant-CaVE} reduces the constant CaVE to the propositional CaVE.
    Definition \ref{corollary:grounded-CaVE} reduces the grounded CaVE to the constant CaVE.
    To have a predicate version of CaVE in Definition \ref{theorem:CaVE}, we need to show that the predicate CaVE can be reduced to a grounded CaVE without grounding beforehand, only created through resolution.

    % \begin{proof} [Proof of Theorem \ref{theorem:CaVE}]
        For an integrity constraint.
        Atom $B_i$ in Definition \ref{def:constraint-rule} is either a variable emitter or a verifier.
        The constraint handler is a constraint store that collects the truth values and symbols from emitters during the resolution process.
        The ordering of the values enumerated by the resolution does not affect the outcome, since all relations defined in the verifiers are equivalence relations.
        Once the constraint handler has collected sufficient values, the variables are all grounded.
        When all body literals are evaluated as $\top$, the constraint handler returns true.
        When any body literal is evaluated as $\bot$, the constraint handler returns false.
        Hence, the grounded constraint handlers constructed during resolution behave in the same way as the integrity constraints.
        All grounded constraint handlers will be evaluated as long as the underlying resolution search is complete.
    % \end{proof}

    The CaVE algorithm handles integrity constraints using resolution; if the input logic program contains a contradiction, it is unsatisfiable.
    For example, an emitter of the form $p(X)$ with free variable $X$ and the program has the fact $p(X)$.
    This program has a contradiction: a set of facts asserting that $p(X)$ is true, but an integrity constraint stating that $p(X)$ cannot be true.

\section{CaVE Implementation and Examples in stableKanren}
\label{sec:CaVE-implementation}
    We introduce a new syntax, \textit{constrainto}, for stableKanren, allowing us to write integrity constraints as emitters and verifiers as follows.
    \begin{lstlisting}
(constrainto [(e1) (e2) ...] [(v1) (v2) ...])
    \end{lstlisting}
    The constrainto takes in two lists.
    The first is an emitter list to match the emitter during resolution.
    Once the resolution successfully proves a goal function, it looks up the emitter list for a match and passes the values to the verifiers.
    The second is a verifier list to concatenate using $\land$ to form a constraint storage.
    \begin{lstlisting}
(and (v1) (v2) ...)
    \end{lstlisting}
    We use continuation to implement constraint storage and demonstrate its operation in Section \ref{sec:continuation}.
    Section \ref{sec:csp-in-sk} presents three examples of solving combinatorial search problems (CSP) declaratively using constrainto in stableKanren.

\subsection{Continuation is All You Need}
\label{sec:continuation}
    To illustrate how to use continuation to pass values from emitters to verifiers.
    We use the integrity constraint that ensures no queens attack each other on the same row as an example.
    The integrity constraint is written as follows.
    \begin{lstlisting}
(constrainto [(queen x y) (queen u v)] [(= x u) (not (= y v))])
    \end{lstlisting}
    The continuation is the constraint storage of the two verifiers as follows.
    \begin{lstlisting}
(and (= x u) (not (= y v)))
    \end{lstlisting}
    Trying to run this continuation in Scheme raises an unbound variable exception.
    The verifiers are waiting for values from the emitter to proceed with the computation.
    To pass values to a continuation, we wrap anonymous lambdas that have corresponding variables around the continuation as below.
    \begin{lstlisting}
((lambda (u v)
    ((lambda (x y)
        (and (= x u) (not (= y v))))
    1 3))
2 4)
    \end{lstlisting}
    We define a new macro for wrapping so that whenever the emitter generates new values, it wraps them and passes them to the verifier during resolution.
    \begin{lstlisting}
(define-syntax constraint-constructor
  (syntax-rules () [(_ (params ...) (values ...) expr)
                      `((lambda (params ...) expr) values ...)]))
    \end{lstlisting}
    The constraint-constructor macro uses an anonymous lambda to wrap around an expression.
    We apply the constraint-constructor twice to create the same expression in the previous example.
    \begin{lstlisting}
(constraint-constructor (u v) (2 4) 
  ,(constraint-constructor (x y) (1 3)
     (and (= x u) (not (= y v)))))
    \end{lstlisting}
    Now, this constraint storage or the continuation has sufficient values, and a simple \textit{eval} produces the result.

\subsection{Solving Combinatorial Search Problems in stableKanren}
\label{sec:csp-in-sk}
    This section demonstrates a general declarative approach to solving combinatorial problems using integrity constraints in extended stableKanren.
    The general steps for solving these problems are: first, model the problem as a graph; second, generate all combinations of nodes, edges, and numbers that are picked or not; and last, prune unwanted solutions using constraints derived from the desired solution's properties.
    We use three examples, the Hamiltonian cycle, the SEND+MORE=MONEY puzzle, and the Knight's tour, to apply the general steps.
    We will see different amounts of numeric computations in these examples.
    The Hamiltonian cycle example involves no numeric computations, only pure negations, equality, and unifications.
    The Knight's tour example involves two numeric computations: plus and minus.
    The SEND+MORE=MONEY involves even more numeric computations.
    All examples are fully relational, meaning that one program can serve multiple purposes.
    Each example contains queries to find answers from scratch, complete the partial answers, and verify the answers.
    % We only show a single query example to demonstrate this relational property.
    The problem instances are a graph and/or a set of numbers.
    We assume the reader can create the instances without issue, so we focus solely on presenting the algorithms.

\subsubsection{Hamiltonian Cycle}
\begin{figure}[h]
    \centering
    \includegraphics[scale=0.15]{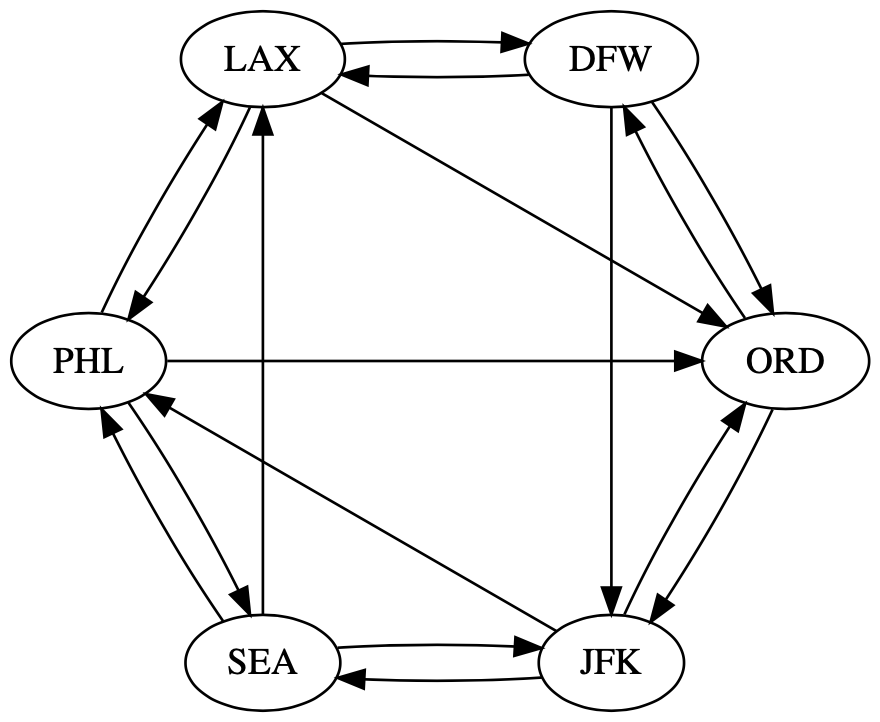}
    \caption{An airline service map}
    \label{fig:airports}
    \end{figure}

    A Hamiltonian cycle is a cycle that visits each vertex exactly once.
    For example, give an airline service map in Figure \ref{fig:airports}, finding a travel plan that visits each airport exactly once.
    The service map is a directed graph, where an arrow means a flight service from one airport to another.
    Firstly, the problem instance is represented using \textit{airport} and \textit{fly} in stableKanren as follows.
    \begin{lstlisting}
(defineo (airport x) (conde [(== x 'DFW)] ...))
(defineo (fly x y) (conde [(== x 'DFW) (== y 'JFK)] ...))
    \end{lstlisting}
    Supplementing the \textit{conde} clauses based on the graph will get a complete graph representation.
    Secondly, each flight service between the two airports can be purchased or not.
    There are 17 flight services, so there are $2^{17}$ possible purchase combinations.
    Lastly, unreasonable purchases are pruned out to get valid travel plans.
    These two steps are written in stableKanren as follows.
    \begin{lstlisting}
(defineo (buy u v) (airport u) (airport v) (fly u v) (noto (free u v)))
(defineo (free u v) (airport u) (airport v) (fly u v) (noto (buy u v)))

(constrainto [(buy u v) (buy x y)] [(eq? x u) (not (eq? v y))])
(constrainto [(buy u v) (buy x y)] [(not (eq? x u)) (eq? v y)])

(defineo (reachable v)
  (conde [(buy 'DFW v)]
         [(fresh (u) (airport u) (buy u v) (reachable u))]))
(constrainto [(airport u) (noto (reachable v))] [(eq? u v)])
    \end{lstlisting}
    The \textit{buy} and \textit{free} generate all possible travel plans, either buying a ticket or not.
    There are three types of invalid travel plans: first, flying out of the same airport more than once; second, flying into the same airport more than once; and third, not flying to an airport at all.
    Therefore, three constraints prune out unreasonable travel plans.
    Unlike the other two examples, no numeric computation is involved in the constraints.

    A set of queries and their corresponding travel plan outputs is presented below.
    \begin{lstlisting}
> (run 1 (q) (fresh (a b c d e f)
               (buy a b) (buy b c) (buy c d) (buy d e) (buy e f) (buy f a)
               (== q `(,a ,b ,c ,d ,e ,f ,a))))
((DFW JFK PHL SEA LAX ORD DFW))
> (run* (q) (buy 'JFK 'PHL) (buy 'PHL 'SEA) (buy 'SEA 'LAX)
             (buy 'LAX 'DFW) (buy 'DFW 'ORD) (buy 'ORD 'JFK))
(_.0)
> (run* (q) (buy 'JFK 'PHL) (buy 'PHL 'SEA) (buy 'SEA 'DFW)
            (buy 'DFW 'LAX) (buy 'LAX 'ORD) (buy 'ORD 'JFK))
()
> (run 1 (q) (fresh (a b d e f)
               (buy a b) (buy b 'DFW) (buy 'DFW d) (buy d e) (buy e f) (buy f a)
               (== q `(,a ,b DFW ,d ,e ,f ,a))))
((JFK ORD DFW LAX PHL SEA JFK))
> (run 1 (q) (fresh (a d e f)
               (buy a 'SEA) (buy 'SEA 'DFW) (buy 'DFW d) (buy d e) (buy e f) (buy f a)
               (== q `(,a SEA DFW ,d ,e ,f ,a))))
()
    \end{lstlisting}
    The first query asks for a travel plan.
    It returns one starting at DFW.
    The second query verifies a travel plan starting at JFK.
    It confirms the travel plan is valid.
    The third query verifies another travel plan starting at JFK.
    It says the plan is infeasible.
    The fourth query concerns DFW as the third stop on the trip.
    It returns a valid travel plan.
    The last query proposes SEA as the second stop and DFW as the third stop.
    It is impossible to meet the requirement.

\subsubsection{Knight's Tour}
    Knight's Tour is given an $m \times n$ chessboard and a knight that visits each position on the chessboard exactly once, creating a Hamiltonian path.
    The minimum viable chessboard for a solution is a $5\times5$ board \cite{Martin:1995:KnightTourNumbers}.
    Firstly, the problem instance is represented using \textit{nums} and \textit{steps} in stableKanren as follows.
    \begin{lstlisting}
(defineo (nums x) (conde [(== x 1)] ... [(== x 5)]))
(defineo (steps x) (conde [(== x 1)] ... [(== x 25)]))
    \end{lstlisting}
    For a $5\times5$ board, there are 25 steps.
    So the domains of \textit{nums} and \textit{steps} are 5 and 25, respectively.
    Secondly, each step $s$ can pick or not pick a position $(x,y)$.
    There are 25 steps and 25 positions, so it has $25^{25}$ combinations.
    Lastly, invalid jumps are pruned to obtain a valid traverse path.
    The algorithm is as follows. 
    \begin{lstlisting}
(defineo (pick s x y) (nums x) (nums y) (steps s) (noto (free s x y)))
(defineo (free s x y) (nums x) (nums y) (steps s) (noto (pick s x y)))

(constrainto [(pick s1 x1 y1) (pick s2 x2 y2)]
             [(= s1 s2) (not (= x1 x2)) (not (= y1 y2))])
(constrainto [(pick s1 x1 y1) (pick s2 x2 y2)]
             [(= x1 x2) (= y1 y2) (not (= s1 s2))])

(defineo (jumped s) (fresh (x y) (pick s x y)))
(constrainto [(steps s1) (noto (jumped s2))] [(= s1 s2)])

(constrainto [(pick s1 x1 y1) (pick s2 x2 y2)]
             [(= s2 (+ s1 1)) (not (directions x1 x2 y1 y2))])
(define (directions x1 x2 y1 y2)
  (or (and (= x2 (+ x1 2)) (= y2 (+ y1 1)))
      (and (= x2 (+ x1 1)) (= y2 (+ y1 2)))
      (and (= x2 (- x1 1)) (= y2 (+ y1 2)))
      (and (= x2 (- x1 2)) (= y2 (+ y1 1)))
      (and (= x2 (- x1 2)) (= y2 (- y1 1)))
      (and (= x2 (- x1 1)) (= y2 (- y1 2)))
      (and (= x2 (+ x1 1)) (= y2 (- y1 2)))
      (and (= x2 (+ x1 2)) (= y2 (- y1 1)))))
    \end{lstlisting}
    The \textit{pick} and \textit{free} generate all combinations of steps and positions, the knight either jumps to position $(x,y)$ on step $s$ or does not jump.
    There are four types of invalid jumps: first, jumping to two positions in the same step; second, jumping to the same position more than one step; third, not jumping to a position at all; and fourth, the jump between two steps violates the knight's legal movement directions.
    Therefore, four constraints prune out unreasonable tour plans.
    Specifically, the \textit{directions} is a Scheme function, and the numeric computations are only carried out in the verifiers to ensure the program is relational.

    A set of queries and their corresponding knight's tour outputs is as follows.
    \begin{lstlisting}
> (run 1 (q)
    (fresh (x1 y1 x2 y2 x3 y3 x4 y4 x5 y5 x6 y6 x7 y7 x8 y8 x9 y9 x10 y10 x11 y11
            x12 y12 x13 y13 x14 y14 x15 y15 x16 y16 x17 y17 x18 y18 x19 y19 x20 y20
            x21 y21 x22 y22 x23 y23 x24 y24 x25 y25)
      (pick 1 x1 y1) (pick 2 x2 y2) (pick 3 x3 y3) (pick 4 x4 y4) (pick 5 x5 y5)
      (pick 6 x6 y6) (pick 7 x7 y7) (pick 8 x8 y8) (pick 9 x9 y9) (pick 10 x10 y10)
      (pick 11 x11 y11) (pick 12 x12 y12) (pick 13 x13 y13) (pick 14 x14 y14)
      (pick 15 x15 y15) (pick 16 x16 y16) (pick 17 x17 y17) (pick 18 x18 y18)
      (pick 19 x19 y19) (pick 20 x20 y20) (pick 21 x21 y21) (pick 22 x22 y22)
      (pick 23 x23 y23) (pick 24 x24 y24) (pick 25 x25 y25)
      (== q `(( ,x1 ,y1 ) ( ,x2 ,y2 ) ( ,x3 ,y3 ) ( ,x4 ,y4 ) ( ,x5 ,y5 )
              ( ,x6 ,y6 ) ( ,x7 ,y7 ) ( ,x8 ,y8 ) ( ,x9 ,y9 ) (,x10 ,y10)
              (,x11 ,y11) (,x12 ,y12) (,x13 ,y13) (,x14 ,y14) (,x15 ,y15)
              (,x16 ,y16) (,x17 ,y17) (,x18 ,y18) (,x19 ,y19) (,x20 ,y20)
              (,x21 ,y21) (,x22 ,y22) (,x23 ,y23) (,x24 ,y24) (,x25 ,y25) ))) )
(((1 1) (2 3) (3 1) (1 2) (2 4) (4 5) (3 3) (5 2) (4 4) (2 5) (1 3) (2 1) (4 2)
  (5 4) (3 5) (1 4) (2 2) (4 1) (5 3) (3 2) (5 1) (4 3) (5 5) (3 4) (1 5)))
    \end{lstlisting}
    The first query finds an answer.
    It returns one valid Knight's tour path starting at $(1, 1)$ \footnote{Without any heuristic and using BFS as the search strategy, it takes around 1,300 seconds to produce an answer on an Intel i9 iMac 2020. It takes around 220 seconds to produce an answer using DFS on the same machine. }.
    % No heuristic, on an Intel based iMac DFS, 1000 seconds, BFS 170 seconds.
    \begin{lstlisting}
> (run 1 (q)
    (fresh (x1 y1 x2 y2 x5 y5 x6 y6 x10 y10 x11 y11 x15 y15 x16 y16
            x17 y17 x18 y18 x22 y22 x23 y23 x24 y24 x25 y25)
      (pick 1 x1 y1) (pick 2 x2 y2) (pick 3 1 5) (pick 4 3 4) (pick 5 x5 y5)
      (pick 6 x6 y6) (pick 7 4 2) (pick 8 5 4) (pick 9 3 5) (pick 10 x10 y10)
      (pick 11 x11 y11) (pick 12 4 1) (pick 13 3 3) (pick 14 2 5)
      (pick 15 x15 y15) (pick 16 x16 y16) (pick 17 x17 y17) (pick 18 x18 y18)
      (pick 19 2 4) (pick 20 4 5) (pick 21 5 3) (pick 22 x22 y22)
      (pick 23 x23 y23) (pick 24 x24 y24) (pick 25 x25 y25)
      (== q `((,x1 ,y1) (,x2 ,y2) (1 5) (3 4) (,x5 ,y5)
              (,x6 ,y6) (4 2) (5 4) (3 5) (,x10 ,y10)
              (,x11 ,y11) (4 1) (3 3) (2 5) (,x15 ,y15)
              (,x16 ,y16) (,x17 ,y17) (,x18 ,y18) (2 4) (4 5)
              (5 3) (,x22 ,y22) (,x23 ,y23) (,x24 ,y24) (,x25 ,y25) ))) )
(((1 1) (2 3) (1 5) (3 4) (1 3) (2 1) (4 2) (5 4) (3 5) (1 4) (2 2) (4 1) (3 3)
  (2 5) (4 4) (5 2) (3 1) (1 2) (2 4) (4 5) (5 3) (3 2) (5 1) (4 3) (5 5)))
    \end{lstlisting}
    The second query gives a few blocks of partial answers for the 3rd, 4th, 7th, 8th, 9th, 12th, 13th, 14th, 19th, 20th, and 21st steps.
    It completes the unsolved steps in the partial answers.

    \begin{lstlisting}
> (run 1 (q)
    (fresh (x1 y1 x2 y2 x5 y5 x6 y6 x10 y10 x11 y11 x12 y12 x16 y16
            x17 y17 x18 y18 x22 y22 x23 y23 x24 y24 x25 y25)
      (pick 1 x1 y1) (pick 2 x2 y2) (pick 3 1 5) (pick 4 3 4) (pick 5 x5 y5)
      (pick 6 x6 y6) (pick 7 4 2) (pick 8 5 4) (pick 9 3 5) (pick 10 x10 y10)
      (pick 11 x11 y11) (pick 12 x12 y12) (pick 13 4 1) (pick 14 3 3)
      (pick 15 2 5) (pick 16 x16 y16) (pick 17 x17 y17) (pick 18 x18 y18)
      (pick 19 2 4) (pick 20 4 5) (pick 21 5 3) (pick 22 x22 y22)
      (pick 23 x23 y23) (pick 24 x24 y24) (pick 25 x25 y25)
      (== q `(( ,x1 ,y1 ) ( ,x2 ,y2 ) ( 1 5 ) ( 3 4 ) ( ,x5 ,y5 )
              ( ,x6 ,y6 ) ( 4 2 ) ( 5 4 ) ( 3 5 ) (,x10 ,y10)
              (,x11 ,y11) (,x12 ,y12) (4 1) (3 3) (2 5)
              (,x16 ,y16) (,x17 ,y17) (,x18 ,y18) (2 4) (4 5)
              (5 3) (,x22 ,y22) (,x23 ,y23) (,x24 ,y24) (,x25 ,y25) ))) )
()
    \end{lstlisting}
    The third query also gives a partial answer, but the 12th to 14th steps in the previous query are shifted to the 13th to 15th steps.
    This change causes the query to return no results.

\subsubsection{SEND+MORE=MONEY}
    The SEND+MORE=MONEY puzzle requires choosing different single-digit numbers (0-9) for each letter (SENDMORY) to satisfy the equation.
    An additional restriction is that there are no leading zeros in the numbers, so S and M can not be 0.
    Firstly, the problem instances include \textit{letters} and \textit{values}, the \textit{letters} will have 8 symbols (SENDMORY), and the \textit{values} will have 10 symbols (0-9).
    Secondly, each letter can be assigned a value or not.
    There are 8 letters and 10 digits, so it has $8^{10}$ combinations.
    Lastly, unsatisfied assignments are pruned to obtain the answer.
    The algorithm is shown as follows.
    \begin{lstlisting}
(defineo (assign l v) (letters l) (values v) (noto (free l v)))
(defineo (free l v) (letters l) (values v) (noto (assign l v)))

(constrainto [(assign l1 v1) (assign l2 v2)]
             [(eq? l1 l2) (not (= v1 v2))])
(constrainto [(assign l1 v1) (assign l2 v2)]
             [(not (eq? l1 l2)) (= v1 v2)])

(defineo (assigned l)
  (fresh (v) (letters l) (values v) (assign l v)))
(constrainto [(letters l1) (noto (assigned l2))] [(eq? l1 l2)])

(constrainto
   [(assign 's s) (assign 'e e) (assign 'n n) (assign 'd d)
    (assign 'm m) (assign 'o o) (assign 'r r) (assign 'y y)]
   [(not (= (+ (* s 1000) (* e 100) (* n 10) (* d 1)
               (* m 1000) (* o 100) (* r 10) (* e 1))
    (+ (* m 10000) (* o 1000) (* n 100) (* e 10) (* y 1))))])
    \end{lstlisting}
    The \textit{assign} and \textit{free} generate all possible combinations of the letters and values.
    % The variables $l$ and $v$ in the \textit{assign} get values through \textit{letters} and \textit{values} as we have explained in Section \ref{sec:loops}.
    There are four types of invalid assignment: first, one letter takes more than one value; second, the same value is assigned to more than one letter; third, not assigning a value to a letter at all; and fourth, the assignment did not satisfy the SEND+MORE=MONEY equation.
    Therefore, four constraints prune out unreasonable assignments.
    Specifically, the last constraint performs numerical computations to verify the SEND+MORE=MONEY equation.

    % A query uses the \textit{run} interface to produce an answer to the puzzle as follows.
    A set of queries and their corresponding outputs is as follows.
    \begin{lstlisting}
> (run 1 (q)
    (fresh (s e n d m o r y)
      (assign 's s) (assign 'e e) (assign 'n n) (assign 'd d)
      (assign 'm m) (assign 'o o) (assign 'r r) (assign 'y y)
      (== q `(,s ,e ,n ,d ,m ,o ,r ,y))))
((9 5 6 7 1 0 8 2))
> (run 1 (q) (assign 'm 2))
()
> (run 1 (q)
    (fresh (s d m o r y)
      (assign 's s) (assign 'e 5) (assign 'n 6) (assign 'd d)
      (assign 'm m) (assign 'o o) (assign 'r r) (assign 'y y)
      (== q `(,s 5 6 ,d ,m ,o ,r ,y))))
((9 5 6 7 1 0 8 2))
    \end{lstlisting}
    The first query produces an answer to the puzzle.
    The second query verifies that the assignment is invalid.
    The third query completes a partial answer. 

\section{Conclusion and Future Work}
    In conclusion, this paper presents a new constraint storage algorithm, CaVE, designed to handle integrity constraints with resolution.
    We show that CaVE is equivalent to the integrity constraints, provided all constraints are equality relations; the completeness of CaVE depends on the underlying resolution, as CaVE operates on the side of resolution.
    We implement CaVE in stableKanren and add a new syntax \textit{constrainto} to stableKanren.
    In stableKanren, we have seen macros in Scheme that simplify logic-program translation and provide support for stable model semantics \cite{Guo:2023:PPDP-stableKanren}.
    In this paper, the continuation from functional programming reduces the effort required for implementing constraint storage in CaVE.
    The extended stableKanren encodes declarative solutions to various combinatorial search problems that also preserve the relational property since the non-unification computations are performed only in the verifiers.

    There are many improvements to be made.
    Firstly, the integrity constraints in this paper are hard constraints; once a constraint is violated, the partial solution is discarded.
    For some problems, it is acceptable to violate constraints with penalties to get a non-optimal solution.
    So, a new resolution-based algorithm for handling soft integrity constraints would be useful.
    Secondly, the verifiers in CaVE check only answers; they can also guide the search process.
    There is no resolution-based heuristic algorithm for boosting integrity constraint-solving performance.
    The Conflict Driven Nogood Learning (CDNL) algorithm is the state-of-the-art algorithm for bottom-up solving \cite{Gebser:2007:conflict-driven-answer-solving} and
    Hanson and Sussman implement a propagation system that contains cells, propagators, dependency-directed backtracking, and can learn the nogoods from contradictions \cite{Hanson:2021:software-design-flexibility}.
    Hence, having a resolution-based CDNL would improve the performance.
    Thirdly, the emitters in CaVE emit safe values that require all variables to be bound, and the constraints are evaluated immediately after all values are collected.
    Adding support for emitting unbound variables and delayed constraint evaluation will enable CaVE to work with definite programs, providing a feature similar to constraint logic programming (CLP).
    Lastly, and more specifically, when the symbols are numbers, CLP techniques can be used for improving solving speed.
    The SEND+MORE=MONEY puzzle can be solved by cKanren in under a second \cite{Alvis:2011:ckanren}.
    Hence, an algorithm that identifies the CLP property in the integrity constraints and reduces the program to CLP would be helpful.

\bibliographystyle{eptcs}
\bibliography{ref}
\end{document}